\newcommand{\f}[1]{Fig.~\ref{#1}}
\newcommand{\eq}[1]{Eq.~(\ref{#1})}

\def\be{\begin{equation}}
\def\ee{\end{equation}}
\def\bea{\begin{eqnarray}}
\def\eea{\end{eqnarray}}
\def\l({\left(}
\def\r){\right)}

\def\arccosh{\rm arccosh}
\def\ds{\displaystyle}
\def\Bm{B_a^{\max}}

  \newcommand{\PRL}{{Phys.\ Rev.\ Lett.~}}
	  \newcommand{\PR}{{Phys.\ Rev.~}}
	  
	  \newcommand{\JAP}{{J.\ Appl.\ Phys.~}}
	  
	  \newcommand{\SUST}{{Supercond.\ Sci.\ Technol.~}}
     	  
	  \newcommand{\PhysC}{{Physica C~}}

\documentstyle[prb,aps,multicol,graphicx,psfig]{revtex}

 \renewcommand{\narrowtext}{\begin{multicols}{2} 
\global\columnwidth20.5pc}
 \renewcommand{\widetext}{\end{multicols} \global\columnwidth42.5pc}

\begin{document}
\title{Magnetostrictive behaviour of thin superconducting disks }
\author{Tom~H.~Johansen$^{1,}$\cite{0} and Daniel~V. Shantsev$^{1,2}$
}
\address{
$^1$Department of Physics, University of Oslo, P. O. Box 1048
Blindern, 0316 Oslo, Norway\\
$^2$A. F. Ioffe Physico-Technical Institute, Polytekhnicheskaya 26,
St.Petersburg 194021, Russia\\
}
\date{\today}
\maketitle

\begin{abstract}
Flux-pinning-induced stress and strain distributions 
in a thin disk superconductor
in a perpendicular magnetic field is analyzed. 
We calculate the body forces, solve the magneto-elastic problem and 
derive formulas for all stress and strain components, including the magnetostriction
$\Delta R/R$. The flux and current density profiles in the disk
are assumed to follow the Bean model.
During a cycle of the applied field 
the maximum tensile stress is found to occur approximately midway
between the maximum field and the remanent state.
An effective relationship between this overall maximum stress and the peak field 
is found.   
\end{abstract}

%\pacs{PACS numbers: 74.76.-w, 74.25.Ha, 74.60.Ge, 68.60.Dv}
%74.76.Bz, 74.60.Jg

%\narrowtext

\section{Introduction}

During recent years top-seeded growth of bulk Y-Ba$_2$Cu$_3$O$y$ and 
(RE)-Ba$_2$Cu$_3$O$y$ 
(RE = rare earth elements) has developed enormously towards production of 
large single-grain superconductors with strong flux pinning \cite{re123-review,NEG-review}.
An impressive critical current density value of 
$J_c = 100$ kA/cm$^2$ at 77 K has now been obtained for bulk 
(Nd-Sm-Gd)-Ba$_2$Cu$_3$O$y$ with 10 mol\% of Gd-211.\cite{NEG-review}
Large-grain superconductors of this 
type can trap very high magnetic fields in the remanent state, 
exceeding the field from conventional permanent magnets by an order of 
magnitude and even more.  Therefore, it is today strongly believed that the 
field-trapping ability of bulk superconductors is useful in many
practical applications like magnetic separators, 
magnetron sputtering systems, motors etc. %\cite{mura,mizutani,ostwald}.  

A major problem of such bulks is that mechanical 
fracturing frequently occurs during the magnetic activation process. 
The strong flux pinning unavoidably leads to large body forces, which 
during descent of the applied field are tensile. 
Due to the irreversible nature of the flux dynamics, the resulting stress 
distribution becomes non-uniform, and can locally
reach the tensile strength of the material so that fatal cracking takes place.
A systematic study of this problem was first made by Ren et al. 
\cite{ren}, who also made model calculations of the internal stress in a 
cylindrical superconductor during the field descent after field-cooling. 
Later the modelling of the magneto-elastic behaviour has been extended 
considerably by including various geometrical cases and magnetization
conditions \cite{thj-prl,thj-rect,thj-cyl,thj-sust,thj-jap,thj-clamped}. 

Common to all previous modelling work is that the superconductor is 
approximated as infinitely long in the direction of the applied field.
Today, one sees that single-grain disks with diameters up to 6 cm 
are grown on routine basis,\cite{mizutani} 
whereas low-gravity environment
has allowed growth of even larger disks - up to 12.7 cm in diameter and 
2 cm in thickness.\cite{murakami}. When the diameter becomes much larger 
than the thickness the infinitely long 
cylinder approximation is known to give a poor description of flux and 
current distributions.
A better approximation can then be a very thin sample in perpendicular field, where  
exact analytical results for flux distribution are also available. They were
used to calculate magnetostriction for thin long strips in Ref.\onlinecite{nab-2}, 
but only for the virgin branch.    
In this work we use the thin 
disk solution, and find all the stresses and strains for the full magnetization cycle.    
We present here results of this model and discuss the
relationship between the peak applied field and the overall maximum tensile 
stress occurring in the disk during pulsed field magnetic activation.

\section{Magneto-Elastic Problem}

Consider a thin superconducting circular disk of radius $R$ and thickness $d$,
where $d \ll R$. The superconductor is placed in a magnetic field, $B_a$, applied
perpendicular to the disk plane, i.e., along the $z$-axis.
When the local field at the edge exceeds the lower critical field, $B_{c1}$,
the vortices start to penetrate into the disk. Due to strong demagnetization 
effects this happens 
already at $B_a \ge \sqrt{d/R}\; B_{c1}$. The presence of pinning centers
will cause the vortices to distribute non-uniformly over the disk volume.
According to the critical state model the quasi-stationary flux
distribution adjusts irreversibly so that the local current density never exceeds
the critical magnitude $J_c$. In analyses of this behaviour a common simplification is to 
ignore the effects of a lower critical field, and also the presence of any
reversible magnetization. Under these conditions analytical results for the field and
current distributions have been derived for the thin disk geometry.\cite{clem}

\begin{figure}
\centerline{\includegraphics[width=7cm]{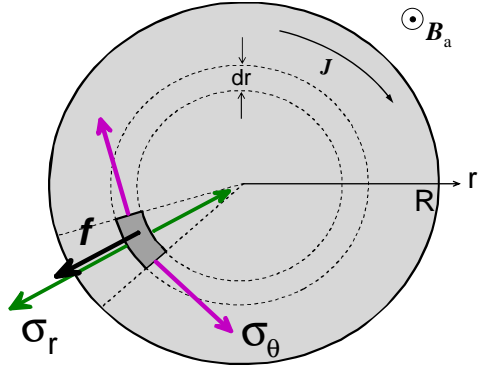}}
\caption{\label{f_1} Internal stresses and body forces in a superconducting disk 
placed in a perpendicular applied magnetic field.}
\end{figure}

The force per unit volume acting on the vortices is given by the Lorentz force
${\bf f} = {\bf J} \times {\bf B}$, where ${\bf J}$ and ${\bf B}$ are 
the local current and flux density, respectively. 
In quasi-static equilibrium this force is balanced by pinning forces, 
and hence, ${\bf f}$ also represents the distribution of body forces transmitted
onto the crystal lattice in a given magnetized state.
The symmetry of the disk geometry implies that the induced current 
 has only an azimuthal component,  
and the force averaged over the disk thickness
 points in the radial direction, see \f{f_1}.

For the thin disk one can therefore write
\be
 f(r) = J(r) B(r),
\label{2}
\ee     
where $J(r)$ is the current density averaged over $d$,
and $B(r)\equiv B_z(z=0)$ is the flux density in the disk's center plane.

Exact analytical results for a thin disk in perpendicular field are available
only for the Bean model, where $J_c$
is assumed field-independent. We adopt this model in the present analysis, and
review here only the results needed for the calculations of the
magneto-elastic behavior.

Starting with a flux free superconductor the application of a magnetic field, $B_a$,
will cause penetration of flux into the disk up to the point $r=a$, where
\be
a = \frac{R}{\cosh (B_a/B_d)} \quad {\rm and} \quad B_d = \mu_0J_c d/2.
\label{3}
\ee
In the annular vortex-filled region the flux density has the strongly non-linear
profile given by
\be
\frac{B(r)}{B_d} = \frac{B_a}{B_d} - \arccosh(R/r) +
\frac{2}{\pi} \int^{\pi/2}_{\arcsin(a/r)} 
\frac{1-\theta \cot \theta}{\sqrt{1-(r/R)^2 \sin\theta}}\, d\theta,
\quad a \le r \le R.
\label{4}
\ee
The central region $r \le a$ has $B=0$, as for the long cylinder.

\begin{figure}
\centerline{\includegraphics[width=9cm]{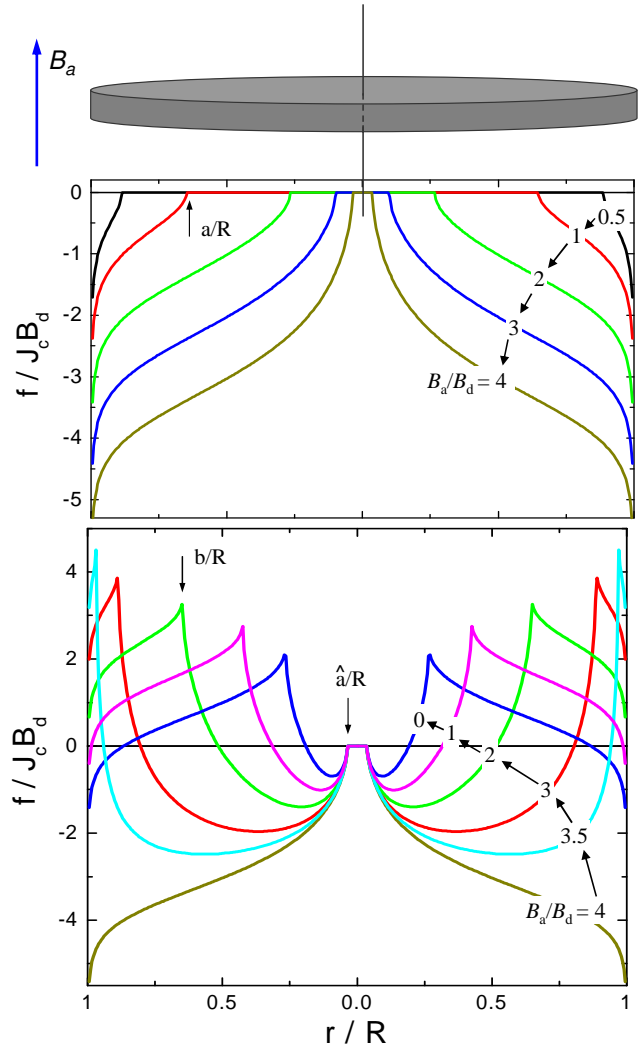}}
\caption{\label{f_2} Body force distribution in a disk during field rise from the 
virgin state (top), and subsequent field descent (bottom). Positive force 
means that it points radially {\em out} from the center.}
\end{figure}

The shielding current is for a thin disk distributed over the entire area
according to      
\be
J(r)/J_c = \left\{ 
\begin{array}{ll}
-1, & a \le r < R, \\
\ds{
- \frac{2}{\pi} \arctan \l( \frac{r}{R} \sqrt{\frac{R^2-a^2}{a^2-r^2}} \r)}, & r \le a. 
\end{array}
\right.
\label{5}
\ee
Note here that $J$ is {\em not} proportional to the flux density gradient $dB/dr$,
thus preventing us from using the relation $fdr=-d(B^2)/2\mu_0$, which
is valid for a long cylinder.
Figure~\ref{f_2} (top) shows graphs of the body force distribution, Eq.(1),
during virgin magnetization of the disk. Since the current has a constant
magnitude where the flux density is non-zero,
the force profile has the same shape as the function $B(r)$.

When the applied field is reduced after reaching a maximum value
$\Bm$, the disk undergoes remagnetization where the front is located at $r=b$, with
\be
b = \frac {R}{\cosh\left[ (\Bm-B_a)/2B_d\right]} .
\label{6}
\ee 
In the inner region, $r<b$, the flux density generated by the maximum field 
is unchanged, whereas in the outer annulus, $b \le r < R$, considerable redistribution
of flux is taking place. The current, on the other hand,
becomes modified in the entire disk area. During field reduction the disk
is therefore divided into three regions, each described by its own
set of functions $J(r)$, $B(r)$ and thereby also $f(r)$.
One has for:\\
Region I (outer); $b \le r <R$;
\bea
\frac{B(r)}{B_d} = \frac{B_a}{B_d} + \arccosh(R/r) +
\frac{2}{\pi} \left[
\int^{\arcsin(b/r)}_{\arcsin(\hat{a}/r)} 
\frac{1-\theta \cot \theta}{\sqrt{1-(r/R)^2 \sin\theta}}\, d\theta -
\int^{\pi/2}_{\arcsin(b/r)} 
\frac{1-\theta \cot \theta}{\sqrt{1-(r/R)^2 \sin\theta}}\, d\theta
\right],\nonumber \\
J(r)/J_c = 1 .
\label{7}
\eea
Region II (middle); $\hat{a} \le r <b$;
\bea
\frac{B(r)}{B_d} = \frac{\Bm}{B_d} - \arccosh(R/r) +
\frac{2}{\pi} \int^{\pi/2}_{\arcsin(\hat{a}/r)} 
\frac{1-\theta \cot \theta}{\sqrt{1-(r/R)^2 \sin\theta}}\, d\theta,\nonumber \\
\frac{J(r)}{J_c} = 
- 1 + \frac{4}{\pi} \arctan \l( \frac{r}{R} \sqrt{\frac{R^2-b^2}{b^2-r^2}} \r)
\label{8}
\eea
Region III (inner); $r<\hat{a}$;
\bea
B(r) = 0,\nonumber \\
\frac{J(r)}{J_c} = 
\frac{2}{\pi} \left[ 
2 \arctan \l( \frac{r}{R} \sqrt{\frac{R^2-b^2}{b^2-r^2}} \r) -
\arctan \l( \frac{r}{R} \sqrt{\frac{R^2-\hat{a}^2}{\hat{a}^2-r^2}} \r)
\right]
\label{9}
\eea

Shown in \f{f_2} (bottom) are plots of $f(r)$ at various stages of field reduction.
A cusped peak in the body force is seen to accompany the remagnetization front 
at $r=b$. After reversing the field sweep the outer part of the
disk immediately starts to experience forces that will create tension. 
When the field has been reduced to zero considerable remanent body forces, a mixture 
of tensile and compressive forces, remain in the disk.

To determine the quantitative magnetostrictive behaviour a separate analysis of
the thin disk elasticity problem is required. 
For simplicity, we assume that
the superconductor is mechanically isotropic, and that the strains are well below the 
fracture limit allowing linear elasticity theory to be applicable. 
As mechanical boundary conditions for the disk we choose a free surface, i.e.,
the normal stresses, $\sigma_r$ and $\sigma_z$ vanish at the surfaces.
Moreover, since the disk is thin one can assume that $\sigma_z=0$
throughout the volume, and that the deformation is described by only a radial 
displacement field, $u(r)$. The strain components are then
\be
e_r = u'(r)\quad {\rm and} \quad e_\theta = u/r,
\label{10}
\ee
which are related to the non-vanishing stresses $\sigma_r$ and $\sigma_\theta$ by
\be
Ee_r = \sigma_r - \nu \sigma_\theta \quad {\rm and} \quad
Ee_\theta = \sigma_\theta - \nu \sigma_r .
\label{11}
\ee
Here $E$ and $\nu$ is the Young's modulus and Poisson's ratio, respectively.
The condition of static equilibrium, see \f{f_1}, is that
\be
\sigma_r'(r) + \frac{\sigma_r-\sigma_\theta}{r} + f = 0,
\label{12}
\ee      
which in terms of the displacement field is expressed as
\be
u''+\frac 1r u' - \frac 1{r^2}\ u + \frac{1-\nu^2}{E} f = 0,
\label{13}
\ee
or 
\be
\frac{d}{dr} \left[ \frac 1r \frac{d(ur)}{dr} 
\right] = - \frac{1-\nu^2}{E} f .
\label{14}
\ee

Integrating twice, and using that $\sigma_r(r=R)=0$, and $u(0)=0$, one
obtains
\be
u(r) = \frac{1-\nu}{2E} r \left[
\left( \frac{1-\nu}{R^2} + \frac{1+\nu}{r^2} 
\r) \int_0^r r'^2 f dr' + \int^R_r \l(
1+\nu+\frac{1-\nu}{R^2} r'^2 \r) f dr'
\right]  .
\label{15}
\ee
It immediately follows that the external dilatation $\Delta R = u(R)$
is given by
\be
\frac{\Delta R}{R} = \frac{1-\nu}{E} \frac{1}{R^2} \int_0^R r^2 f dr.
\label{16}
\ee
The internal strains can be expressed as 
\bea
\begin{array}{l}
e_r(r)\\
e_\theta(r)
\end{array}
= \frac{1-\nu}{2E} \left[
\left( \frac{1-\nu}{R^2} \mp \frac{1+\nu}{r^2} 
\r) \int_0^r r'^2 f dr' + \int^R_r \l(
1+\nu+\frac{1-\nu}{R^2} r'^2 \r) f dr'
\right] ,
\label{17}
\eea
with the upper and lower sign corresponding to $e_r$ and $e_\theta$, respectively.
Similarly, the radial stress and the hoop stress are given by
\bea
\begin{array}{l}
\sigma_r(r)\\
\sigma_\theta(r)
\end{array}
= \frac{1-\nu}{2} \left[
\left( \frac{1}{R^2} \mp \frac{1}{r^2} 
\r) \int_0^r r'^2 f dr' + \int^R_r \l(
\frac{1+\nu}{1-\nu}+\frac{r'^2}{R^2} \r) f dr'
\right] .
\label{18}
\eea
The eqs.(\ref{15}-\ref{18}) are exact expressions from which 
the magneto-elastic response can be calculated once the force
distribution $f(r)$ is known. One can see immediately that the two strains and
the two stresses are equal in the center, i.e., $e_r(0)=e_\theta(0)$ and
$\sigma_r(0)=\sigma_\theta(0)$, independent of the magnetized state.
Notice also that the hoop stress at the rim of the disk is proportional to
the dilatation, $\sigma_\theta(R)=E\Delta R/R$, as it should 
according to \eq{11} and the condition $\sigma_r(R)=0$.

\section{Results and Discussion}

\subsection{Magnetostriction $\Delta R/R$}

The irreversible behaviour of the magnetostriction, $\Delta R/R$, as 
a function of the applied field, $B_a$, is calculated from \eq{16}
using the expressions for $J(r)$ and $B(r)$ that were listed in the 
previous section. The result is shown in \f{f_4} (left),
where hysteresis loops with three different maximum fields are plotted.
The overall shape of the loops is quite similar to that found in the long cylinder 
case.\cite{thj-cyl} When the maximum applied field becomes much higher than $B_d$
a large part of the loop is well represented by two straight lines forming a cross 
centered at $B_a=0$. Asymptotically, this linear behaviour is described by
\be
\frac{\Delta R}{R} = \frac{1-\nu}{E} J_c R B_d \l( 
\pm \frac{B_a}{3B_d} + 7.57 \cdot 10^{-2} \r).
\label{19}
\ee
Therefore, the width of large field ($\Bm\gg B_d$) hysteresis loop
rapidly approaches 
\be
(\Delta R/R)_{\downarrow} - (\Delta R/R)_{\uparrow} = 
2 \frac{1-\nu}{E}\ J_c R B_a . 
\label{20}
\ee
Interestingly, the same simple relation holds also for the long cylinder when
$\Bm>2B_p$, where $B_p=\mu_0J_cR$ is the full penetration
field.\cite{thj-cyl}   
The virgin branches of the loops are not included in \f{f_4} since on 
the chosen scale they are barely resolved.

The corresponding result for thin long strips is given by
Eqs.~16-17 of Ref.\onlinecite{nab-2}. For large $\Bm$ it reduces to our 
\eq{20}, except that the factor $1-\nu$ is missing.     
This factor should appear when the stress from the currents on the far edges 
of the strip is taken into account, as also discussed in Ref.\onlinecite{nab-2}.   

Dilatation loops found experimentally\cite{ikuta-prl,nab-2} can deviate from
the shown \f{f_4} (left) due to a field dependence of the critical current.
If this dependence is rather strong, the loops will narrow down for large fields,
as shown in \f{f_4} (right). These curves were calculated assuming an
exponential dependence of the critical current, $J_c(B)=J_c(0) \exp(-|B|/2B_d)$.
Flux and current distributions cannot be written down explicitly in case of
field-dependent $J_c$,
however they can be found numerically by solving a set of integral equations given
in Ref.~\onlinecite{jcb1}. It allows calculation of magnetization loops, as
explained in Ref.~\onlinecite{jcb2}, where $M$ decreases with $B$ at large fields
in accordance with chosen $J_c(B)$ dependence. A similar procedure was used
to calculate the dilatation loops of \f{f_4} (right).

\begin{figure}
\centerline{\includegraphics[width=8.5cm]{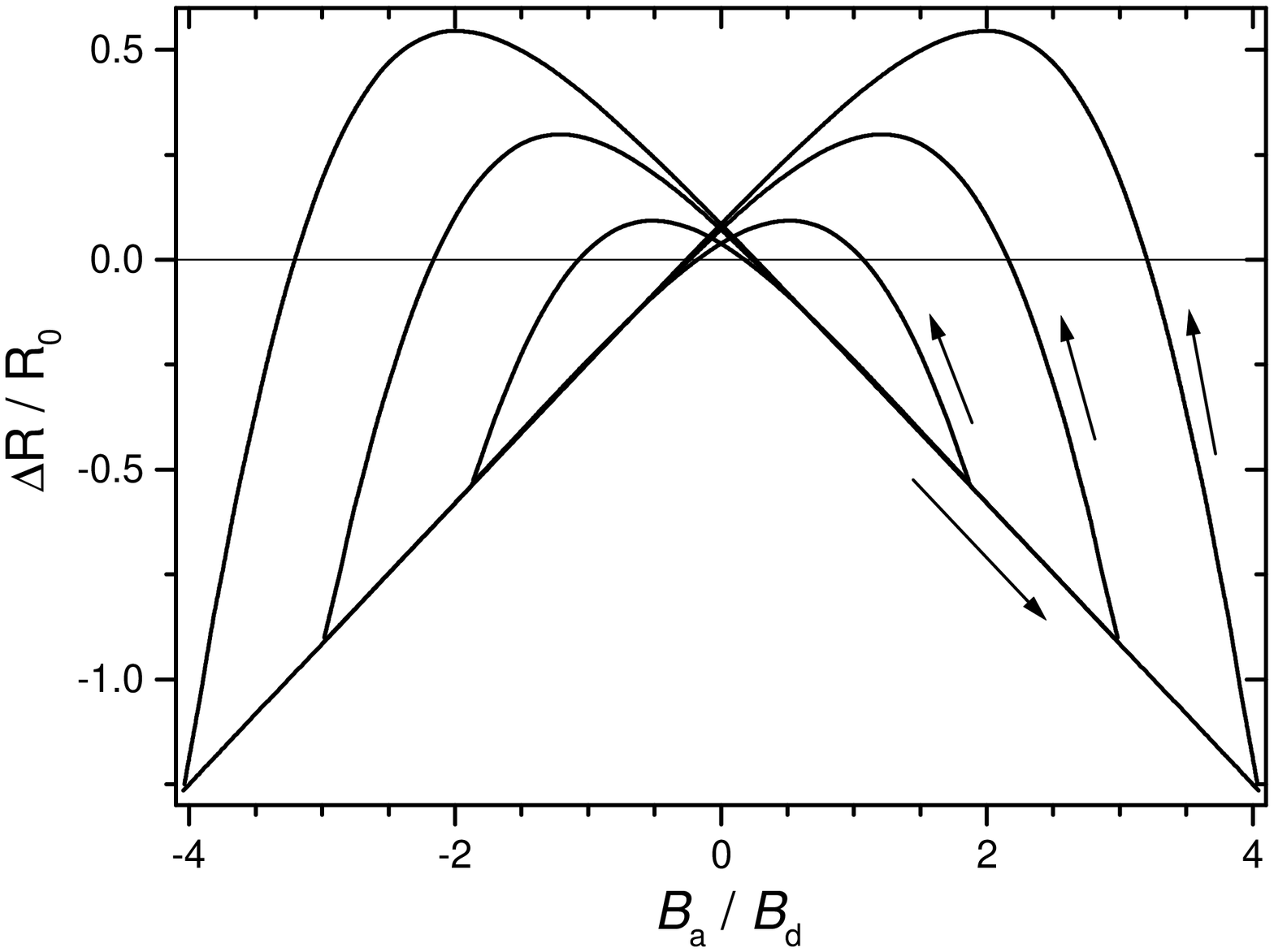}
\includegraphics[width=8.5cm]{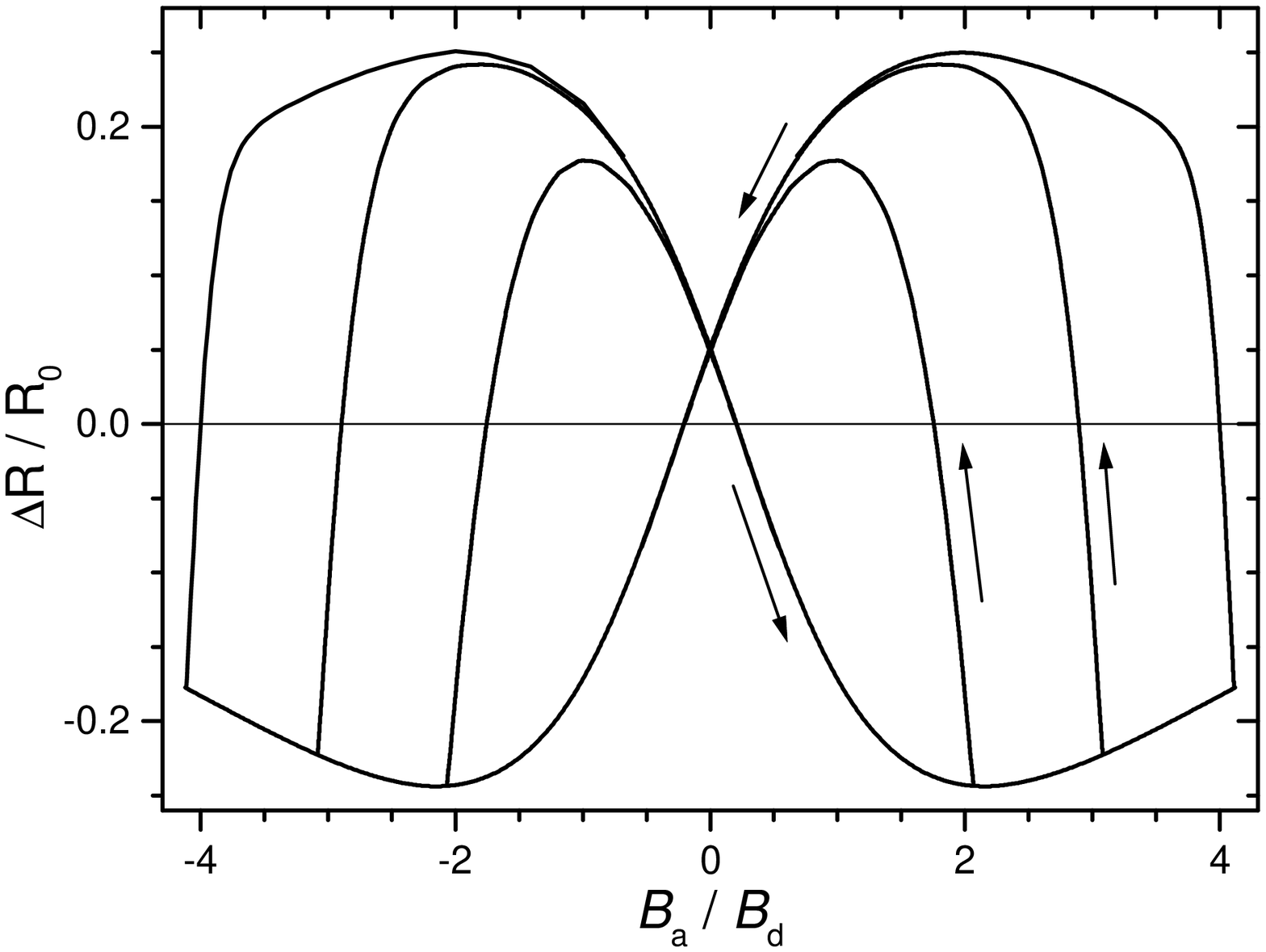}}
\caption{\label{f_4} Dilatation loops for cycles of $B_a$ with a different
maximum field. The dilatation is normalized by $R_0 = (1-\nu)J_c B_d R / E$. 
Left: the Bean model. Right: a field-dependent critical current, 
$J_c(B)=J_c(0) \exp(-|B|/2B_d)$.}
\end{figure}

\subsection{Stress Distributions}

As the applied field increases, the body force points towards the disk centre,
leading to stresses that everywhere are compressive (negative). 
During the ramp up of the field the compressive stress grows monotonically, 
and always with a spatial distribution where the central unpenetrated region
experiences the largest compression. Since these stress values are normally much 
lower than the compressive strength of the superconductor, we focus here instead 
on the descending field stage of the magnetic activation process.

\begin{figure}
\centerline{\includegraphics[width=8cm]{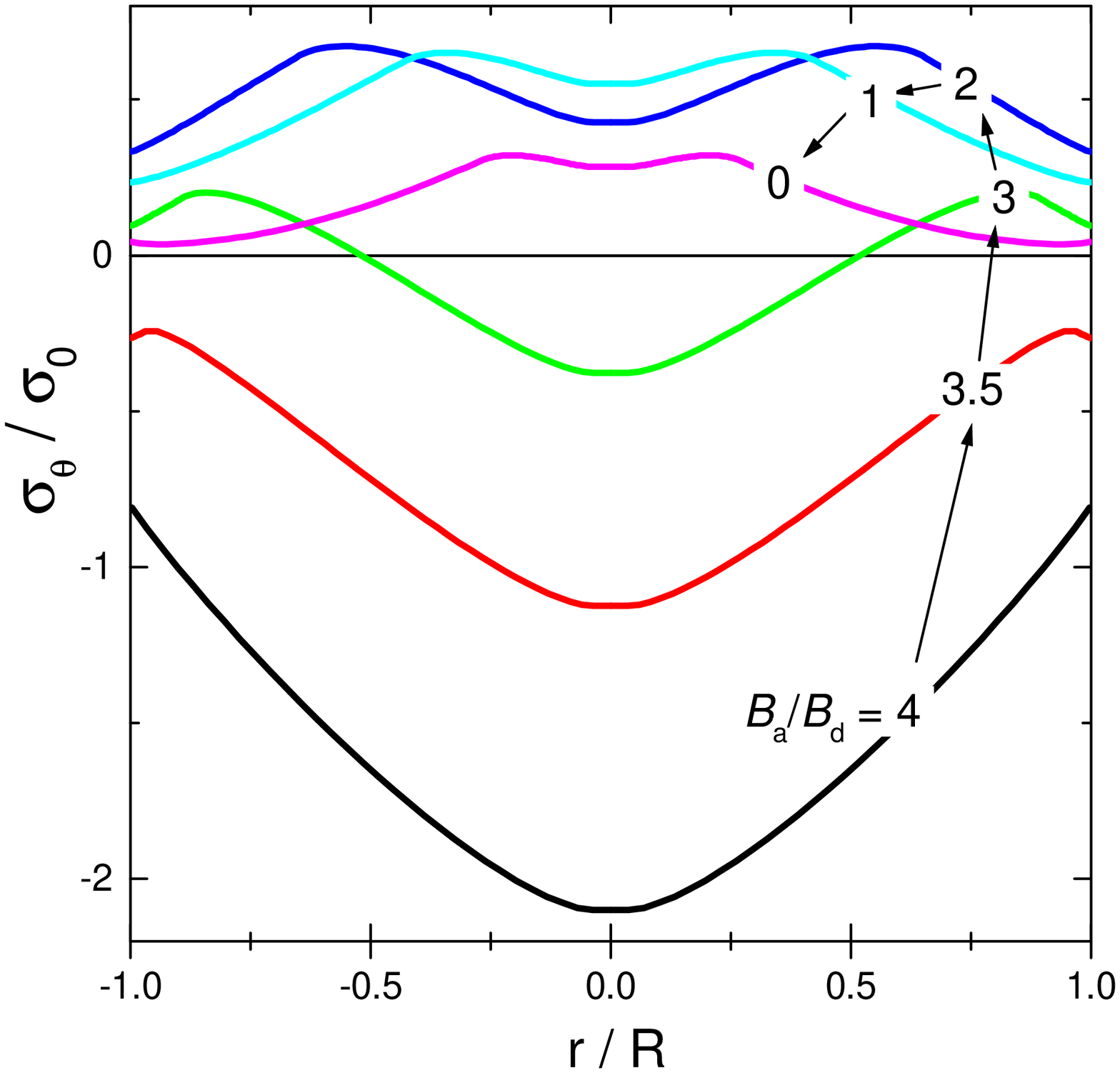}
\includegraphics[width=8cm]{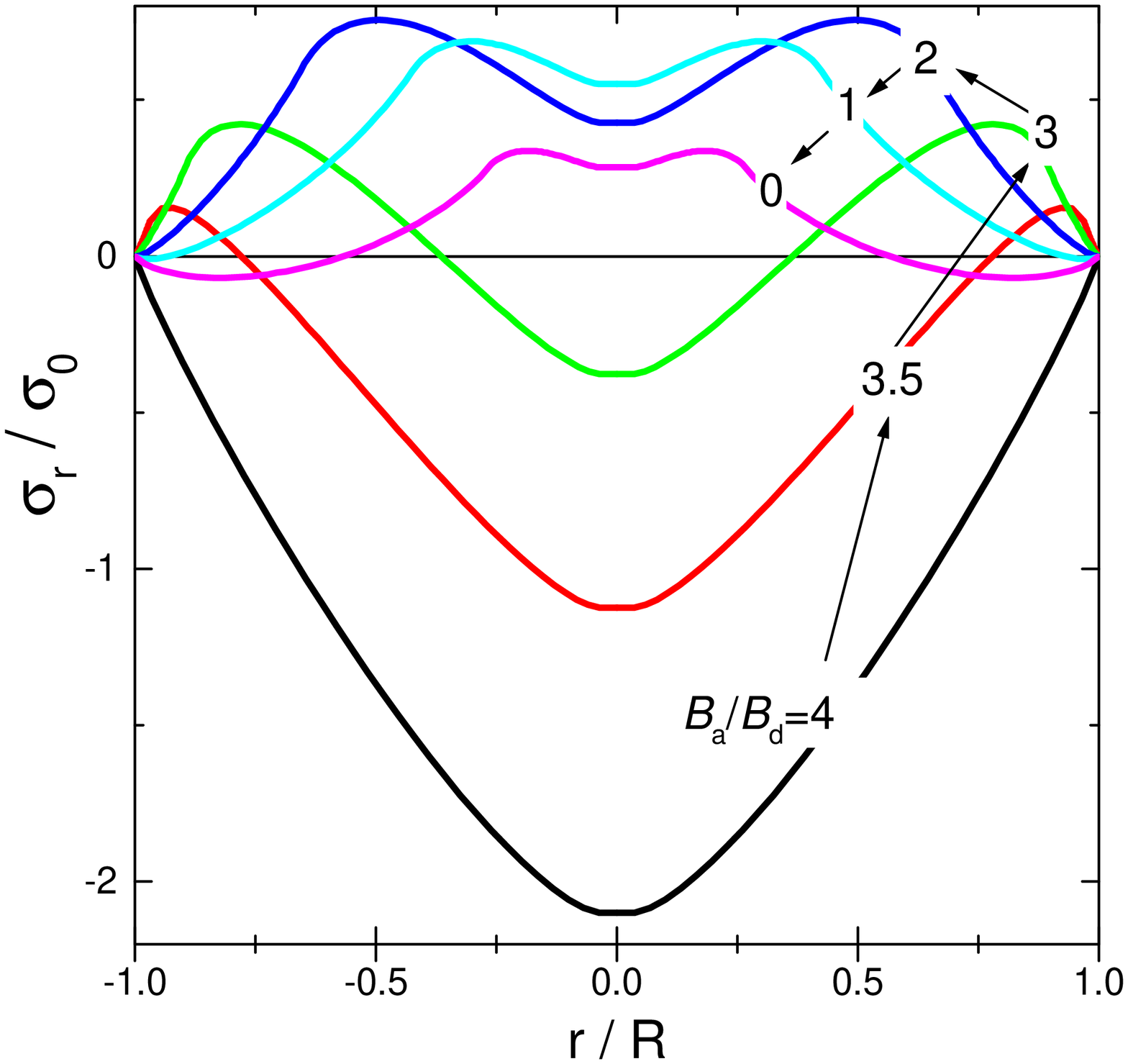}}
\caption{\label{f_3} Distribution of the hoop (left)
and radial (right) stresses in a disk during the field decent after reaching a 
maximal value $B_a=4B_d$. The stress is normalized by $\sigma_0 \equiv B_dJ_cR$
and $\nu=0.3$ was used in the calculations.}
\end{figure}

As an example, let us consider the case where the maximum applied field has
the value $\Bm = 4 B_d$.
When $B_a$ starts to decrease, the stress distribution becomes
more complex since the body force turns expansive in the outer remagnetized region.
The \f{f_3} illustrates how the stresses develop during the field descent.
Very soon a peak of tensile (positive) stress appears near the remagntization front, 
and the peak value grows in height.
At an intermediate field the peak tension reaches an overall maximum, which for 
the radial stress amounts to 0.8$\sigma_0$, where $\sigma_0 = B_dJ_cR$.
The maximum hoop stress is slightly smaller. It is during this part of the 
field descent the disk is most likely to crack, namely the first time the material 
encounters a stress equal to the tensile strength. 
 
\newpage 
\narrowtext
\begin{figure}
\centerline{\includegraphics[width=9cm]{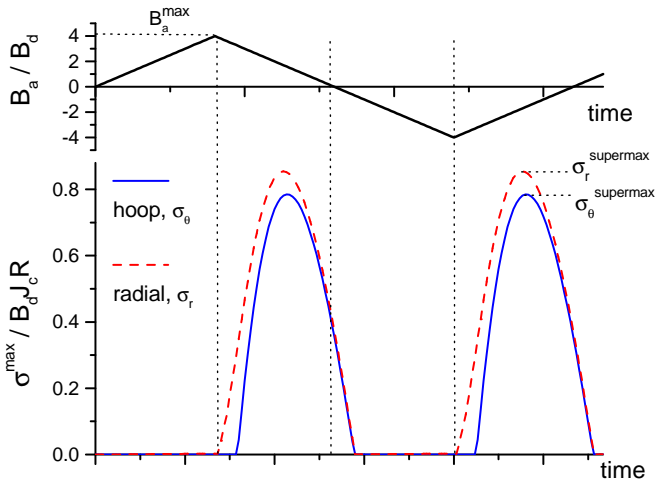}}
\caption{\label{f_time} Time evolution of maximal tensile stress in the disk 
during a full cycle of applied field starting from the virgin state.
}
\end{figure}

\begin{figure}
\centerline{\includegraphics[width=9cm]{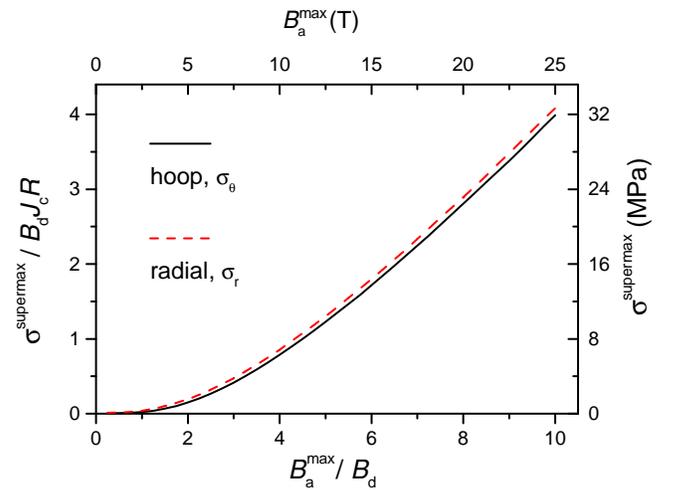}}
\caption{\label{f_maxmax} The overall maximal stress reached during
the full field cycle, as found from \f{f_time}, versus the applied field amplitude.
The dimensional values on the right and top axes are given for a typical
disk with parameters given in the text.  
}
\end{figure}
\widetext

Shown in \f{f_time} is the time evolution of the stresses during one full cycle of
the applied
field. The curves show the maximum tension occurring in the disk.
The parts given zero value correspond to the stages where the stress is everywhere 
compressive. It becomes evident that
 the most 'dangerous' stage of the process is slightly
after midway between the maximum field and the remanent state.
The tension in the remanent state is roughly twice smaller than
the overall maximum value, which we denote as $\sigma^{\rm supermax}$.
The next Figure shows how this $\sigma^{\rm supermax}$ depends
on the maximum applied field in the cycle. 
The dependence is essentially parabolic, and is well fitted by a 
simple formula,
\be
\label{parab}
\frac{\sigma^{\rm supermax}}{B_dJ_cR} = 0.063\ \frac{\Bm}{B_d}+0.035\l(\frac{\Bm}{B_d}\r)^2 
\ee
By substituting the following realistic numbers for disk parameters;
$J_c=10$~kA/cm$^2$, $d=2$~cm, $R=3$~cm, 
one obtains dimensional values for
the peak tension versus the maximal applied field as shown in
the top and right axes of \f{f_maxmax}. This graph represents the essence of
the present work, and may hopefully serve as a useful guide in practical situations.
E.g. it follows that for a disk with tensile strength of 20 MPa,\cite{tomita} 
one has to apply a field less than 18 T to avoid cracking during the pulsed-field
activation.

The results presented here are obtained in the limit of thin disk in perpendicular field.
Combined with the previous results for a long cyliner\cite{thj-cyl}, 
they give two extreme cases between which all real samples 
fall into. Unfortunately, the analytical treatment for samples with finite thickness is not 
possible, and numerical results are not available. Therefore, we may only hope that the key
quantities, like magnetostriction and maximal tensile stress, show a smooth interpolation
behavior between the two extreme cases. Additional reason for this hope is a very smooth
interpolation behavior for magnetization and the penetration field that have
already been calculated numerically for arbitrary thickness.\cite{Brandt96}
For example, the normalized $M(H)$ curves for all thickness/radius ratios differ
by not more than 3\% from some average $M(H)$.
It should be mentioned that our magnetostriction plots 
for long cylinder\cite{thj-cyl} and thin disk, \f{f_4}, also show a remarkable similarity.   

In conclusion,
we solve the magneto-elastic problem and 
find stress, strain and magnetostriction in a thin disk superconductor
for the full cycle of applied field.  
The maximum tensile stress occurs approximately midway
between the maximum field and the remanent state, see \f{f_time},
and has quasi-parabolic dependence on the peak field, see \eq{parab}. 
Qualitative behavior of the stress and strain components, and the normalized
magnetostriction loop is similar to the case of long cylinder.
 
\subsection{Acknowledgements}

The work was financially supported by the Norwegian 
Research Council and NorFa.

\end{document}